\documentclass[prl,10pt,twocolumn,showpacs]{revtex4}
\usepackage[dvips]{epsfig}
\usepackage{float}
\usepackage{amsfonts}
\usepackage{amsmath}
\usepackage{amssymb}
\usepackage{enumerate}

\begin{document}
\title{Probing the dispersive and spatial properties of planar photonic crystal waveguide modes via highly efficient coupling from optical fiber tapers}

\author{Paul E. Barclay}
\email{pbarclay@caltech.edu}
\author{Kartik Srinivasan}
\author{Matthew Borselli}
\author{Oskar Painter}
\affiliation{Department of Applied Physics, California Institute of Technology, Pasadena, California 91125}
\date{\today}
\begin{abstract}
The demonstration of an optical fiber based probe for efficiently exciting the waveguide modes of high-index contrast planar photonic crystal (PC) slabs is presented.  Utilizing the dispersion of the PC, fiber taper waveguides formed from standard silica single-mode optical fibers are used to evanescently couple light into the guided modes of a patterned silicon membrane.  A coupling efficiency of approximately $95\%$ is obtained between the fiber taper and a PC waveguide mode suitably designed for integration with a previously studied ultra-small mode volume high-Q PC resonant cavity\cite{ref:Srinivasan3}.  The micron-scale lateral extent and dispersion of the fiber taper is also used as a near-field spatial and spectral probe to study the profile and dispersion of PC waveguide modes.  The mode selectivity of this wafer-scale probing technique, together with its high efficiency, suggests that it will be useful in future quantum and non-linear optics experiments employing planar PCs.       

\end{abstract}

\maketitle

\setcounter{page}{1}

Photonic crystals (PCs), formed by periodically modulating the refractive index of a bulk host material, have attracted great interest in recent years owing to the control they provide of the spatial and dispersive properties of light within them\cite{ref:Yablonovitch}.  Two examples of PC structures which employ this control to significantly modify the optical properties of their host material are waveguides\cite{ref:Johnson4,ref:Loncar2} and  defect cavities\cite{ref:Painter3}.  PC waveguides can be designed to exploit modified group velocity\cite{ref:Notomi2}, increased field intensity, and so-called ``generalized phase-matching''\cite{ref:Cowan1} to enhance optical nonlinear behavior.  Point defects within PCs can in turn be used to form optical cavities which support localized resonances with simultaneously high-Q and small effective mode volume.  Such resonant cavities formed in high-index contrast PC dielectric slabs are particularly interesting for future chip-based cavity QED experiments\cite{ref:Mabuchi} and for quantum optical sources such as triggered single photon emitters\cite{ref:Michler,ref:Santori}, where the extremely large electric field strength per photon of the cavity mode can be used to strengthen the coherent coupling to an atomic or quantum dot effective two-level system or in the ``bad cavity'' limit to channel spontaneous emission more efficiently into the cavity mode\cite{ref:Gerard2}. 
 
While several initial applications of planar PCs have been demonstrated, such as extremely small lasers\cite{ref:Painter3} and micro-optical add-drop filters\cite{ref:Noda2}, a significant barrier to further development of useful photonic crystal devices has been the difficulty in coupling light into and out of the PC.  The small mode profiles, a result of the high refractive index of semiconductor materials from which high-index contrast PCs are typically created, and the complex optical phase within periodic structures make it difficult to use conventional coupling methods from optical fibers or free-space beams\cite{ref:Notomi}.  In this Letter we demonstrate that by using the dispersive properties inherent to planar PCs, one may evanescently couple light in an efficient manner from silica optical fiber tapers\cite{ref:Bulmer,ref:Knight,ref:Spillane2} into a high-index contrast planar PC waveguide\cite{ref:Obrien2,ref:Barclay2,ref:Barclay3}.  As efficient coupling is obtained to a photonic crystal waveguide (PCWG) mode designed for integration with a recently demonstrated high-Q PC cavity\cite{ref:Srinivasan3}, the combined fiber taper-PCWG-PC cavity system promises to provide efficient optical access to the ultra-small mode volumes of PC resonant cavities.  In addition, by utilizing the micron-scale lateral size and dispersion of the tapered region of the fiber, this coupling technique is shown to be useful for mapping the bandstructure and spatial profile of PCWG modes.     

\begin{figure}[ht]
\begin{center}
\epsfig{figure=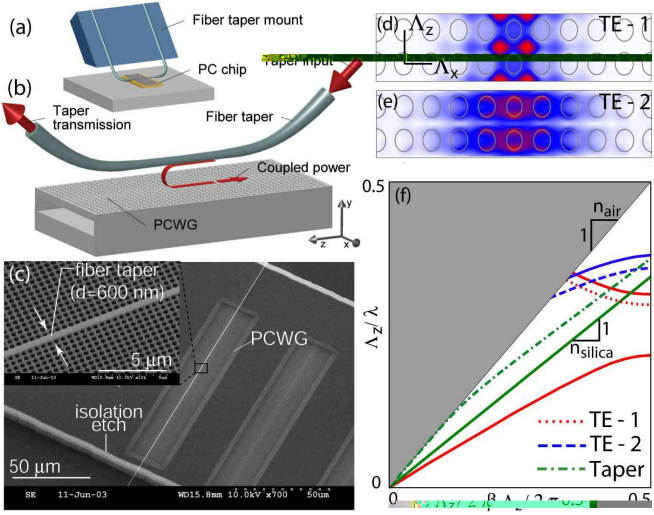, width=1.0\linewidth}
\caption{(a-b) Schematic of the evanescent coupling scheme showing (a) the geometry of the fiber taper mount and (b) the position of the fiber taper relative to the PCWG during coupling.  (c) Scanning electron microscope (SEM) images showing a fabricated PCWG with a fiber taper positioned at its center.   Magnetic field profiles of the (d) TE-1 and (e) TE-2 PCWG modes. (f) Two-dimensional effective-index ($n_{\text{eff}}=2.64$) bandstructure calculation of a compressed square lattice  ($\Lambda_{\text{x}}/\Lambda_{\text{z}}\sim0.8$) of air holes with radius $r = 0.35\Lambda_{\text{x}}$.  Only modes with TE-like polarization and with dominant Fourier components in the $\Gamma$-X ($\hat{z}$) direction are shown.  Donor-type defect PCWG modes are shown as dashed and dotted lines.  The dot-dashed line shows the dispersion of the fundamental mode of a fiber with diameter $d = 3\Lambda_z \sim 1.5$ $\mu$m.   
}
\label{fig:Couple_Scheme}
\end{center}
\end{figure}

The optical coupling scheme used in this work is shown schematically in Figure \ref{fig:Couple_Scheme}(a,b).  An optical fiber taper\cite{ref:Birks1}, formed by heating and stretching a standard single-mode silica fiber, is placed above and parallel to a PCWG.  The fiber diameter changes continuously along the length of the fiber taper, reaching a minimum diameter on the order of the wavelength of light.  Light that is initially launched into the core-guided fundamental mode of the optical fiber is adiabatically converted in the taper region of the fiber into the fundamental air-guided mode, allowing the evanescent tail of the optical field to interact with the PCWG; coupling occurs to \emph{phase-matched} PCWG modes which share a similar momentum component down the waveguide at the frequency of interest.  In order to probe the planar PC chip, the fiber taper is mounted in a ``u''-shaped configuration such that it forms an approximately 10 mm straight segment at its midsection (Fig.\ \ref{fig:Couple_Scheme}(a)).  The mounted fiber taper is then placed on a vertical axis stage driven by a DC motor with 50 nm encoder resolution, allowing the taper to be accurately positioned at varying heights above individual PCWG devices (Fig.\ \ref{fig:Couple_Scheme}(b,c)).  The overall insertion loss in the mounted fiber tapers used here was less than $10\%$ (more controlled formation of fiber tapers has shown that insertion loss of $ < 0.1$ dB is possible \cite{ref:Birks1}).

\begin{figure}[b]
\begin{center}
\epsfig{figure=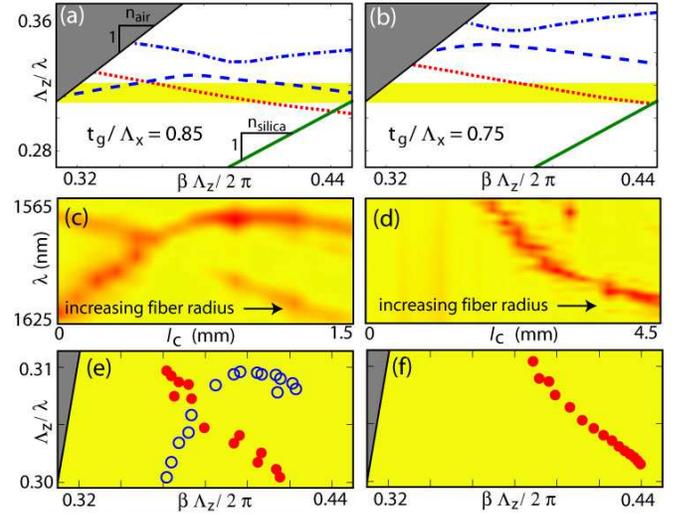, width=1.0\linewidth}
\caption{3D FDTD calculated dispersion of the TE-1 (dotted line), TE-2 (dashed line), and TM-1 (dot-dashed line) modes for the (a) un-thinned ($t_{\text{g}}=340$ nm), and (b) thinned ($t_{\text{g}}=300$ nm) graded lattice PCWG membrane structure ($n_{\text{Si}} = 3.4$).   Measured transmission through the fiber taper as a function of wavelength and position along the tapered fiber for (c) un-thinned sample and (d) thinned sample (different tapers were used for the thinned and un-thinned samples, so the transmission versus $l_{\text{c}}$ data cannot be compared directly).  Transmission minimum  (phase-matched point) for each mode in the (e) un-thinned and (f) thinned sample as a function of propagation constant.  In (a-b), the lightly shaded regions correspond to the tuning range of the laser source used.}
\label{fig:Band_Calc_Measured}
\end{center}
\end{figure} 

The PCWGs studied here were formed on a silicon-on-insulator (SOI) wafer, with a $340$ nm thick silicon (Si) waveguiding layer on top of a silicon dioxide (SiO$_{2}$) underclad layer of $2$ $\mu$m.  Fabrication of the PCWG consisted of electron-beam lithography to pattern an e-beam resist mask, a sulfur hexafluoride reactive ion etch (ICP-RIE) to transfer the PCWG lattice of air holes through the Si waveguiding layer, and a final hydrofluoric acid wet etch to selectively remove the underlying SiO$_{2}$ layer, freeing the Si waveguiding layer and creating a free-standing membrane (Fig.\ \ref{fig:Couple_Scheme}(c)).  A linear array of waveguides was formed on the Si chip, each waveguide of length $200$ lattice constants ($\sim 100$ $\mu$m) with different lattice spacings and nominal hole radius (filling fraction).  An additional isolation etch, a few microns in depth, was performed to remove material surrounding the linear array of PCWGs (Fig.\ \ref{fig:Couple_Scheme}(c)).  Along with the curvature of the fiber taper loop, this ensured that the fiber taper mode would only evanescently interact with the PCWG region. 

Although the evanesent coupling scheme is not specific to a given defect geometry or lattice, the waveguide in this work was chosen to consist of a linear defect along the $\Gamma$-$X$ direction of a compressed square lattice, formed by introducing a lateral grade in the hole radius (Figs.\ \ref{fig:Couple_Scheme}(c-e)).  The choice of grading and lattice compression was motivated by previous work studying the design and integration of waveguides and ultra-small mode volume, high-Q resonant cavities.  This type of PC waveguide supports a mode (the TE-1 mode described below) which has a transverse overlap factor of greater than $98\%$ with that of the high-Q resonant cavity studied in Ref. \cite{ref:Srinivasan3}, and initial finite-difference time-domain (FDTD) simulations of the coupled waveguide-cavity system indicate that efficient loading of the high-Q cavity can be obtained\cite{ref:Barclay2}.  Figure \ref{fig:Couple_Scheme}(f) shows an approximate bandstructure of TE-like (electric field predominantly in the plane of the slab) modes\cite{ref:note_odd_even_modes} of the host compressed square lattice PC slab whose dominant Fourier components are in the direction of the waveguide, and which will consequently couple most strongly with the fundamental fiber taper mode\cite{ref:Barclay2}.  Superimposed upon this bandstructure are the important \emph{donor}-type defect waveguide modes: (i) a defect mode with negative group velocity which comes off of the conduction band edge (labelled TE-1), and (ii) a defect mode with positive group velocity which comes off of a second order (in the vertical direction) valence band-edge (labelled TE-2).  A typical fiber taper dispersion curve is also shown, lying between the air and silica light lines.  

In Figure \ref{fig:Band_Calc_Measured}(a-b), three-dimensional (3D) FDTD simulations were used to accurately calculate the PCWG bandstructure in the regions where phase-matching to the fiber taper mode occurs.  The FDTD-calculated in-plane magnetic field profiles of the TE-1 and TE-2 PCWG modes (taken in the mid-plane of the dielectric slab) near their respective phase-matching points are shown in Fig.\ \ref{fig:Couple_Scheme}(d-e).  Although one can couple to either of the TE-1 or TE-2 modes\cite{ref:Barclay3}, the TE-1 mode is of primary interest here because of its fundamental nature in the vertical direction and its similar properties to that of the high-Q cavity mode of Ref.\ \cite{ref:Srinivasan3}.  Furthermore, coupling to the TE-1 mode is of greater general interest since its phase matching with the silica taper is mediated by the dispersion of the PC, unlike the TE-2 mode which phase matches thanks to its higher order nature in the vertical direction, a mechanism not unique to periodic structures.  The physical scale of the longitudinal ($\Lambda_{\text{z}}$) and transverse ($\Lambda_{\text{x}}$) lattice constants of the PCWG are determined by the phase-matching requirement between the desired TE-1 PCWG mode and the fundamental fiber taper mode, which from the 3D FDTD bandstructure occurs for $\Lambda_{\text{z}} = 500\text{nm}$ ($\Lambda_{\text{x}} = 0.8\Lambda_{\text{z}}=400\text{nm}$) near the operating wavelength of $1600$ nm.  

In order to probe the bandstructure of the PCWG a scanning external cavity laser with a $1565$-$1625$ nm wavelength range was connected to the fiber taper input.  The transmitted power through the fiber taper section and the back-reflected power into the fiber from the taper-PCWG coupling section were both monitored.  By varying the position along the fiber taper of the interaction region between the PCWG and taper (as measured by $l_{\text{c}}$, the distance from the fiber taper diameter minimum), the diameter ($d$) of the fiber taper, and hence the propagation constant ($\beta$) of the fiber taper mode interacting with the PCWG mode, could be tuned.  Tuning from just below the air light-line ($d=0.6$ $\mu$m, $n_{\text{eff}}=\beta c / \omega \sim 1.05$), to just above the silica light-line ($d=4.0$ $\mu$m, $n_{\text{eff}}\sim 1.40$) was possible.  Figure \ref{fig:Band_Calc_Measured}(c) shows the taper transmission as a function of wavelength and sample position ($l_{\text{c}}$) when the taper is centered above the PCWG at a height $g\sim700$ nm from the PCWG surface.  Resonances corresponding to both the TE-1 counter- and TE-2 co-propagating modes can be identified.  SEM measurements of the taper diameter as a function of position ($l_{\text{c}}$) were used to calculate the propagation constant of each resonance, allowing the PCWG modes' dispersion to be plotted (Fig.\ \ref{fig:Band_Calc_Measured}(e)).  The measured bandstructure is in close agreement with the FDTD calculated dispersion of Fig.\ \ref{fig:Band_Calc_Measured}(a), replicating both the negative group velocity of the TE-1 mode and the anti-crossing behaviour of the odd vertical parity TE-2 and TM-1 modes. Figures \ref{fig:Band_Calc_Measured}(d,f) show analogous data obtained by probing the sample after the Si slab thickness $t_g$ has been thinned to better isolate the TE-1 mode in $\omega-\beta$ space.  As predicted by the FDTD simulation (Fig.\ \ref{fig:Band_Calc_Measured}(b)), the TE-1 mode is seen to shift slightly higher in frequency due to the sample thinning whereas the higher-order TE-2 mode shifts more quickly with slab thickness and is effectively ``frozen'' out of the laser scan range.  

\begin{figure}[b]
\begin{center}
\epsfig{figure=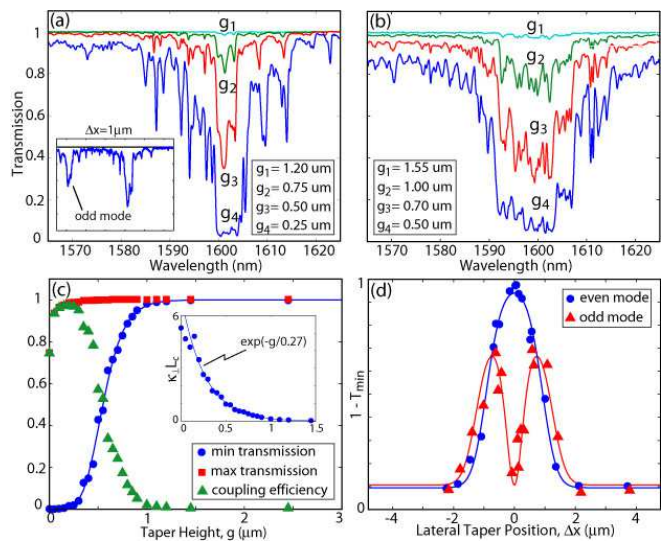, width=1.0\linewidth}
\caption{Coupling characteristics from the fundamental fiber taper mode to the TE-1 PCWG mode of the thinned sample. Transmission versus wavelength for a (a) $1.9\mu$m and (b) $1.0$ $\mu$m diameter taper coupling region for varying taper-PCWG gap, $g$.   (c)  $T_{\text{min}}$, $T_{\text{max}}$ (off-resonance), and $\Gamma$ versus $g$ with a $1.9\mu$m diameter fiber taper coupling region ((inset) effective coupling strength ($\kappa_{\perp}L_{\text{c}}$) versus $g$). (d) $1-T_{\text{min}}$ versus lateral position ($\Delta x$) of the $1.0$ $\mu$m diameter fiber taper relative to the center of the PCWG ($g=400$ nm).  Transmission in (a-d) has been normalized to the transmission through the fiber-taper in absence of the PCWG.  In (d), $1-T_{\text{min}}$ does not go to zero due to non-resonant scattering loss ($\sim 10\%$) for the small taper diameter and taper-PCWG gap used in this measurement.}
\label{fig:Data}
\end{center}
\end{figure}

The dependence of the guided wave coupling on coupling strength and the overall efficiency of the coupling process were studied by varying the gap between the fiber taper and PCWG.  Figure \ref{fig:Data}(a) shows the transmission through the fiber taper, with the coupling region occurring at a taper diameter of $d = 1.9$ $\mu\text{m}$, for varying taper heights above the thinned PCWG\cite{ref:taper_height_note}.  Figure \ref{fig:Data}(b) shows the same measurement, but with the coupling region occurring at a taper diameter of $d = 1.0$ $\mu\text{m}$ and for a PCWG with slightly smaller nominal hole size.  In both cases, the central resonance feature occurring around $\lambda=1600$ nm is the coupling to the TE-1 mode of the PCWG.   In the $d = 1.9$ $\mu\text{m}$ case the minimum transmission ($T_{\text{min}}$) drops below $1\%$ at the phase matching wavelength of $1602$ nm, and recovers to close to unity off resonance.  The taper-PCWG gap ($g$) dependence of $T_{\text{min}}$ shown in Figure \ref{fig:Data}(c) has the hyperbolic functional form $\tanh^2(\kappa_{\perp} L_{\text{c}})$ characteristic of contra-directional coupling\cite{ref:Yeh2} for a coupling strength ($\kappa_{\perp} L_{\text{c}}$) which depends exponentially on $g$ (Fig.\ \ref{fig:Data}(c) inset).  The degree of ideality of this coupling scheme ($\Gamma$) can be estimated by subtracting the off-resonance scattering loss (the scattering loss is expected to be broadband) from $1-T_{\text{min}}$, and is plotted as a function of $g$ in Fig.\ \ref{fig:Data}(c) (green triangles).  The maximum $\Gamma$ occurs at a gap height of $g=250$ nm and is $> 95\%$ in this case.  This near unity coupling efficiency was further confirmed by studying the back-reflected power in the fiber taper.  The back-reflected power ($R$) results from input power that is coupled into the PCWG, reflected by the waveguide termination (in this case an abrupt end to the PC lattice of holes), and coupled back into the backward propagating fiber taper mode.  The maximum in the back-reflected power was measured to be $R_{\text{max}}=15\%$ of the input power at the phase-matched wavelength where $T_{\text{min}}<1\%$, and an estimate of the TE-1 modal reflectivity ($r^2$) from the Fabry-Perot fringes within the transmission spectra of Figs.\ \ref{fig:Data}(a-b) is $10-20\%$.  These measurements are consistent with the near unity taper-PCWG coupling measured from the taper transmission as they imply $\Gamma = \sqrt{R_{\text{max}}/r^2}>0.87$.  

As illustrated by the difference in the transmission curves of Fig.\ \ref{fig:Data}(a) and \ref{fig:Data}(b), the coupler bandwidth is a strong function of $l_{\text{c}}$, with a bandwidth of $20$ nm for coupling with small diameter taper regions ($d\sim 1.0$ $\mu$m) and a bandwidth of less than $10$ nm for coupling to regions of large taper diameter ($d\sim 1.9$ $\mu$m).  This effect has two main contributions: the variation of the TE-1 PCWG mode group velocity at different points in the bandstructure ($n_{\text{g}} = c\,\delta\beta/\delta\omega\sim 4$ - $6$), and the variation in the taper diameter along the length of the 100 $\mu$m PCWG.  As the propagation constant of the fundamental mode in the nominally $1.9$ $\mu$m diameter taper region varies less with changes in fiber diameter ($\delta \beta/\delta d \sim 0.084\,\omega/c$ $\mu\text{m}^{-1}$) compared to the  nominally $1.0$ $\mu$m diameter taper region ($\delta\beta/\delta d \sim 0.36\,\omega/c$ $\mu\text{m}^{-1}$), the coupling bandwidth is increased for smaller tapers.

In addition to probing the bandstructure of the PCWG modes, the micron-scale lateral extent of the fiber taper may also be used as a near-field probe of the localized nature of the PCWG modes.  Fig.\ \ref{fig:Data}(d) shows the coupling dependence of the TE-1 PCWG mode as a function of lateral displacement of the taper from the center of the PCWG ($\Delta x$).  The full-width at half-maximum (FWHM) of $1-T_{\text{min}}$ versus $\Delta x$ (for $1.0$ $\mu$m diameter taper coupling region) was measured to be $\sim 1.84$ $\mu$m, in close agreement with the value ($2.08$ $\mu$m) obtained using a simple coupled mode theory\cite{ref:Barclay2} incorporating the fields of the FDTD calculated TE-1 mode and the analytically determined fiber mode.    Further evidence of the local nature of the taper probe is given by the emergence of a second resonance at a wavelength of $1575$ nm when the taper is displaced laterally (see Fig.\ \ref{fig:Data}(a) inset).  This corresponds to the odd (about $\Delta x = 0$) counterpart to the TE-1 mode, which is only excited by the even fundamental fiber mode if the waveguide mirror symmetry is broken, through fiber misalignment in this case. 

To summarize, evanescent coupling from an optical fiber taper to a highly confined donor-type photonic crystal waveguide mode has been demonstrated with $\sim95\%$ efficiency and over bandwidths as large as $1.3\%$ of the center wavelength ($20$ nm at $1.6$ $\mu$m wavelength).  This coupling technique is not only extremely efficient, but allows for the rapid optical probing of the spatial and dispersive properties of photonic crystal devices throughout the two-dimensional plane of a micro-chip.   It is envisioned that this preliminary demonstration of a wafer scale optical probe for high-index contrast PCs will open up further experimentation in such areas as cavity QED  and novel photon sources where efficient injection and collection of light is paramount.     

KS would like to thank the Hertz foundation and MB the Moore foundation for financial support. 

\bibliographystyle{science} 
\bibliography{./PBG}

\begin{thebibliography}{10}
\bibitem{ref:note_odd_even_modes}
We choose to focus on the TE-like modes due to their larger dispersion and
  bandgap for connected PC lattices. Experimentally this is accomplished by
  adjusting the polarization of the input fiber taper mode.
\bibitem{ref:taper_height_note}
The relative height of the fiber taper was measured from the encoder of the
  drive stage. A reference point (zero gap) for determining the absolute
  taper-PC gap was provided by bringing the taper into contact with the PCWG.
\end{thebibliography}

\end{document}